# What Einstein meant when he said "God does not play dice ..."

*Vasant Natarajan*

**We analyze Einstein's views on God and religion, and his views on Quantum Mechanics.**

One of Albert Einstein's most famous statements is "God does not play dice with the universe". The common interpretation of this statement contains two myths (or perhaps misunderstandings) that I wish to correct in this article.

The first is that his use of the word 'God' implies that he was a religious person who believed in the existence of God. Nothing could be further from the truth; indeed, Einstein can be described more accurately as an outright atheist. Although his early upbringing was in a highly religious Jewish environment, he soon realized that many of the things described in the Old Testament were not consistent with physical laws. His great contributions to physics came from his belief in precise mathematical laws that govern the natural world. This rational approach is antithetical to the common religious notion of a supernatural God with powers that can overcome natural laws.

We can go as far as saying that, deep down, every person (and certainly every practising scientist) must have this rational streak. You cannot do good science if you do not believe in fundamental immutable laws that govern Nature. Tomorrow, if your computer breaks down, you know it is because some part of the system failed. You call a technician hoping he/she will find out what is wrong and fix it; you certainly don't pray to a God or go to a temple to get it fixed (though you might pray to God that the technician comes quickly!). It is interesting that we are born with this rational bent of mind; in fact, our very survival in the natural world depends on forming a rational picture of what we see, with no room for supernatural or magical events. Experiments have

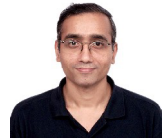

**Vasant Natarajan is at the Department of Physics, Indian Institute of Science. His research interests are in high-resolution laser spectroscopy, optical frequency measurement, and use of laser-cooled atoms to search for time-reversal violation in the laws of physics. He serves on the Editorial board of *Resonance* and contributes regularly to the journal. He considers himself an Ekalavya-type shishya to his guru Einstein. A picture of Einstein hangs above his chair.**

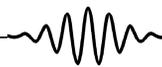

Human infants will get perturbed by magical events that do not conform to their rational model of the world.

shown that human infants below the age of one, well before they are even able to talk, will get perturbed by magical events that do not conform to their rational model of the world (not falling when you go over the edge of a bed, for example). It is only later that we become mature enough to be able to enjoy magic shows by consciously suspending our rational belief during the magician's performance.

So what did Einstein really mean by the word 'God' in his statement? Einstein of course believed in mathematical laws of nature, so his idea of a God was at best someone who formulated the laws and then left the universe alone to evolve according to these laws. He saw the hand of God in the precise nature of physical laws, in their mathematical beauty and elegance, and in their simplicity. To him, the very fact that there were natural laws that the human mind could discover was evidence of a God, not a God who superseded these laws but one who created them. Thus his use of the word God is to be interpreted as the existence of natural laws of great mathematical beauty, whatever form they might take.

Which brings us to the second part of Einstein's statement, the part about not playing dice. This relates to Einstein's reaction to the part of Nature described by *Quantum Mechanics*, which is undoubtedly one of the pillars of modern physics. He felt that natural laws could not be like the throw of dice, with inherent randomness or probability. But this is exactly what Quantum Mechanics tells us – that at the fundamental level Nature is inherently random, codified in Heisenberg's famous Uncertainty Principle. Thus, the second misunderstanding about Einstein's statement is that his opposition to Quantum Mechanics was the raving of an old man, a man well beyond his prime who did not understand the new physics. Well, we will see below why this is all a myth.

Einstein saw the hand of God in the precise nature of physical laws, in their mathematical beauty and elegance, and in their simplicity.

Einstein's great contributions to physics started in his *Annus Mirabilis*, the year 1905 whose centenary was recently celebrated as the World Year of Physics. In that year, Einstein published six seminal papers that revolutionized our understanding of the physical universe in three different directions, and he was all of 26 years old! The papers dealt with (i) the 'light-quantum' or the photon concept and an explanation of the photoelectric effect, (ii) the theory and explanation of Brownian motion, and (iii) the Special Theory of Relativity, a radically new view of space and time. Einstein himself regarded only the first paper as truly revolutionary since it was the second major step (after Max Planck's work) in the development of quantum theory, whereas the Special Theory of Relativity belonged to the older classical theory. In addition, in the same year Einstein discovered the equivalence of mass and energy, encapsulated in perhaps the most famous equation of all: $E = mc^2$.

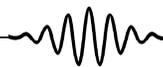

Over the next decade (1905–15), Einstein used his understanding of the 'new' quantum hypothesis to make fundamental contributions to almost every area of physics where the idea could be applied, to the specific heat of solids for example. But he was also quietly working on extending the Special Theory of Relativity to a more generalized theory that would encompass a broader class of transformations between observers. He finally succeeded in 1915, when he published his General Theory of Relativity, a theory of unsurpassed beauty that explained gravitation as arising simply out of the geometry (or curvature) of spacetime. He also showed how the force of gravity had to go beyond the simple but successful theory of Newton. In particular, the force of gravity did not entail Newton's idea of instantaneous action-at-a-distance (e.g. the assumption that the gravitational force of the Sun is felt instantaneously on Earth), but propagated at the speed of light as required by any relativistically correct theory. But this theory was entirely *classical*, there was nothing quantum about it.

The above summary of Einstein's contributions shows two important things about his work: (i) that he made fundamental contributions to our understanding of *quantum theory*, so that he, if any one, was qualified to judge the nature of this theory; and (ii) that, in formulating the General Theory of Relativity, he took our classical ideas beyond what any one had done before. Einstein

In the year 1905 Einstein published six seminal papers that revolutionized our understanding of the physical universe in three different directions, and he was all of 26 years old!

General Theory of Relativity is a theory of unsurpassed beauty that explained gravitation as arising simply out of the geometry (or curvature) of spacetime.

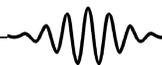

> Einstein was sure that the formalism of the General Theory of Relativity as propounded in 1915 was just a preliminary version that would be extended in due course of time.

spent the remainder of his life in a quest, albeit futile, for an even more generalized theory of relativity; a unified field theory that would geometrize all the forces of Nature and not just gravity. He was sure that the formalism of the General Theory of Relativity as propounded in 1915 was just a preliminary version that would be extended in due course of time. But he was unable to complete this in his lifetime. Here is what one of his biographers, Abraham Pais[1], has to say about his later work – if Einstein had stopped working in 1915, the world of physics would not have lost much. By the way, the same Pais also has the audacity to call the momentous contribution of the great Indian physicist, Satyendra Nath Bose, in formulating his Bose statistics as a lucky "shot in the dark."

[1] Abraham Pais, *Subtle is the Lord: The Science and the Life of Albert Einstein*, by Oxford University Press.

So, did Einstein's contributions to physics end in 1915, did he stop being a part of mainstream physics, was he so completely off the mark that his efforts were doomed before he even started? *Au contraire*. We will now see why his great mind chose to work on the alternate approach to physical theory based on General Relativity, and the basis for his lifelong opposition to Quantum Mechanics.

There were several unique and unprecedented features of General Relativity, three of which were particularly appealing to Einstein:

1. With the equations of General Relativity, Einstein found that space and time were no longer just a passive stage on which particles performed their acts, but were active members in the performance. Thus, the geometric structure of spacetime was determined by the matter in it, and of course the matter responded to this geometry and was constrained by its structure. The fact that space and time were now a part of the equations was unprecedented and its importance is beautifully expressed by Einstein himself: "*It is contrary to the mode of thinking in science to conceive of a thing ... which acts itself, but which cannot be acted upon.*"

> Space and time were no longer just a passive stage on which particles performed their acts, but were active members in the performance.

2. This was the first theory in physics that was *nonlinear*. In

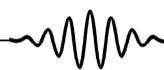

other words, the gravitational field acted upon itself. An important consequence of this was that the equation of motion was contained in the field equations themselves. One did not have separate equations for the interactions between matter and for the response of matter to these interactions. By contrast, a linear theory like Maxwell's theory of electromagnetism could only describe the electromagnetic field interactions between charged matter. The response or inertial manifestation was contained in separate equations of motion given by Newton's three laws of motion that we learn in high school.

3. For the first time in physics, a theory predicted that the inertia of a body (its property of staying at rest or uniform motion unless acted upon by a force, contained in Newton's first law) depended on its surroundings. Much before Einstein, the philosopher-scientist Ernst Mach had the great idea that perhaps the inertia of a body is a consequence of its interactions with the rest of the universe. That is, the distant stars which define the 'inertial coordinates' also determine the inertia of the system. The equations of General Relativity showed that the inertia of a system increases when it is placed in the vicinity of other heavy masses. Inertia was no longer some inherent 'God-given' property of a system, but was at least partly determined by the environment. Einstein's hope was that he would find a fully unified field theory which would show that all of inertia (and not just part of it) was due to interactions with the environment, in keeping with Mach's idea.

We are now in a position to understand Einstein's opposition to Quantum Mechanics. This was not the knee-jerk reaction of a person unqualified to speak about physics, but the considered opinion of an eminent scientist based on what he felt were several undesirable features of the theory. First, he was averse to the idea of randomness as a fundamental feature of any theory. He believed that randomness could appear as some form of statistical behaviour but could not be a part of the law, just like a pack of cards that is shuffled according to deterministic laws still shows a random arrangement. But this was not its only undesirable

> The equation of motion was contained in the field equations themselves.

> The equations of General Relativity showed that the inertia of a system increases when it is placed in the vicinity of other heavy masses.

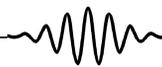

> There were thus three features of Quantum Mechanics that Einstein disapproved of – it was probabilistic, nonlocal, and linear.

feature. The theory was also inherently *nonlocal*, or had a kind of Newtonian action-at-a-distance built into it, but relativity had taught us that all interactions had to propagate at a finite speed. In a landmark paper published in 1935 (twenty years after when Pais considers him finished!), Einstein highlighted the nonlocal and incomplete nature of Quantum Mechanics by proposing the famous Einstein–Podolsky–Rosen (EPR) paradox. The EPR paradox was a *gedanken* or thought experiment that brought out these undesirable features of Quantum Mechanics, and we all know that Einstein was a past master at thought experiments. Finally, of course, Einstein was opposed to the linear formalism of Quantum Mechanics as an approach to understanding Nature because we have seen above that only a nonlinear theory can contain the equations of motion.

There were thus three features of Quantum Mechanics that Einstein disapproved of – it was probabilistic, nonlocal, and linear. Despite this opposition, Einstein realized that it was a successful theory within its domain of applicability. He believed that a future unified field theory would have to reproduce the results of Quantum Mechanics, perhaps as a linear approximation to a deeper nonlinear theory. This was similar to how the relativistic gravitational field of General Relativity (with a finite propagation speed of the gravitational force) led to Newton's law of gravitation (with its action-at-a-distance force) in the nonrelativistic limit. But Einstein was convinced that Quantum Mechanics was not the correct approach to deducing the fundamental laws of physics.

Today, 50 years after his death, the mainstream of physics does not take Einstein's approach seriously. The popular notion is that he was unreasonably opposed to the highly successful Quantum Mechanics. While I have tried to correct this misconception by presenting Einstein's cogent reasoning behind his stand, only time (and perhaps future brilliant scientists who take up his approach) will tell us if he was justified. Let us not forget that Newton's theory of gravity was enormously successful until Einstein came along. We await the next Einstein.

**Suggested Reading**

[1] Vasant Natarajan, Einstein as Armchair Detective: The Case of Stimulated Radiation, *Resonance*, Vol.6, No.6, pp.28–42, 2001.
[2] Vasant Natarajan, V Balakrishnan and N Mukunda, Einstein's Miraculous Year, *Resonance*, Vol.10, No.3, pp.35–56, 2005.
[3] Special Issues on Einstein, *Resonance*, March and April, Vol.5, Nos.3&4, 2000.

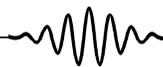

*Postscript: As I finish this article, there is a news item that a letter from Einstein to philosopher Eric Gutkind went on sale in a London auction house. The letter, handwritten in German one year before his death, described belief in God as "childish superstition," and ridiculed the belief that Jews are "the chosen people." The letter sold to someone with "a passion for theoretical physics," for US$ 404000, 25 times the pre-sale estimate. QED.*

*Address for Correspondence*
Vasant Natarajan
Indian Institute of Science
Bangalore 560 012, India.
Email:
vasant@physics.iisc.ernet.in

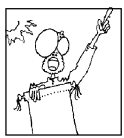

# Information and Announcements

## Symposium on Genomics, Model Organisms and Disease

### National Centre for Biological Sciences (NCBS)
### October 1–2, 2008

The symposium is being organized as a satellite event to the 13th International Human Genome Meeting, being held in India this year (Sept 2008). The focus of this satellite symposium is to highlight genomic approaches in model organisms and how such approaches are furthering understanding of complex diseases. The range of model organisms includes *Drosophila*, C. *elegans*, zebrafish, mouse and humans. This is a unique meeting that brings people using genomic approaches that span RNAi screen, small molecule screens and microarrays together with researchers and doctors who study animal models of human disease.

For more information, contact: www.ncbs.res.in
E-mail: reenas@ncbs.res.in

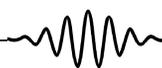